\documentclass[twocolumn,aps,prl,showpacs]{revtex4}
\usepackage[dvips]{graphicx}
\begin{document}
\title{Theory of Bi$_2$Sr$_2$CaCu$_2$O$_{8+\delta}$ Cross-Whisker Josephson Junctions}
\author{Richard A. Klemm}
\email{rklemm@mpipks-dresden.mpg.de}
\affiliation{Max-Planck-Institut f{\"u}r Physik komplexer Systeme,
N{\"o}thnitzer
 Stra{\ss}e 38, D-01187 Dresden, Germany}
\date{\today}
\begin{abstract}
Takano {\it et al.} [Phys. Rev. B {\bf 65}, 140513 (2002) and unpublished]
made  Josephson junctions from single crystal  whiskers of
Bi$_2$Sr$_2$CaCu$_2$O$_{8+\delta}$ crossed an angle $\phi_0$ about the
$c$-axis.  From the mesa structures that formed at the
cross-whisker interface, they inferred a critical current density
$J_c(\phi_0)$.
  Like the single crystal results of Li {\it et
al.} [Phys. Rev. Lett. {\bf 83}, 4160 (1999)], we show that the
whisker data are unlikely to result from a
predominantly $d$-wave order parameter.   However, unlike the single crystals, these
results, if correct, require the whisker $c$-axis transport to be
coherent.
\end{abstract}
\vskip0pt\vskip0pt
\maketitle

Recently, there have been a number of phase-sensitive experiments relevant to the orbital
symmetry of the superconducting order parameter (OP) in the high transition
temperature $T_c$ superconductor Bi$_2$Sr$_2$CaCu$_2$O$_{8+\delta}$
(Bi2212).\cite{tsuei,moessle,kawayama,li,zhu,takano2,takano3,takanosjp,mos2002,yurgens}
 It was claimed that the tricrystal experiment demonstrated a
dominant $d_{x^2-y^2}$-wave OP component in Bi2212 at low temperature $T$ for
both underdoped and overdoped samples.\cite{tsuei}  However, the Pb/Bi2212
$c$-axis Josephson junction experiments demonstrated in many samples that at
least a  small
$s$-wave component was present for $T$ below the $T_c$ of Pb (or Nb).\cite{moessle,kawayama}
In the bicrystal $c$-axis  twist experiment,\cite{li} a dominant $s$-wave OP
for $T\le T_c$ was claimed.\cite{li,bks}  The superb quality of the
junctions was supported by extensive experimental and simulation
analyses including  high resolution transmission electron microscopy
(HRTEM) studies,\cite{zhu} demonstrating that the junctions were atomically clean
over tens of $\mu$m along the junction direction.  The  twist angle 
$\phi_0$
independence of the $c$-axis Josephson critical current density $J_c$ across the
twist junction for $T$ just below $T_c$ was interpreted in terms of
a dominant $s$-wave OP  for all $T\le T_c$,\cite{li,bks} in apparent contradiction to the
results of the tricrystal experiment.\cite{tsuei} However, these experiments
would be compatible if Bi2212 were mostly $s$-wave in the bulk and
$d$-wave on the surface.\cite{mueller}

Single crystal Bi2212 
consists of a stack of intrinsic Josephson junctions, and low-$T$ measurements of the
critical current $I_c$ and the normal resistance $R_n$ across a single
$c$-axis junction led to $I_cR_n$ values $\approx 1/3$ the
Ambegaokar-Baratoff (AB)
result.\cite{yurgens,irie,ab} 
 This is consistent with  non-metallic and incoherent $c$-axis transport in
Bi2212,
\cite{mcguire,tajima,lang} and an  $s$-wave OP.\cite{ab}
Upon intercalation with HgBr$_2$, mesa studies revealed that    
$I_c$ and $R_n$ respectively decreased and increased by two orders of magnitude, but their product $I_cR_n$ remained about 1/3 of the AB
value.\cite{yurgens}  For incoherent transport,
$I_c$ for an $s$-wave  or $d$-wave OP is respectively proportional  to the
$s$-wave  ($1/\tau_s$) or $d$-wave ($1/\tau_d$) interlayer scattering rate,
but  $R_n\propto\tau_s$ for both OP's.  Hence, the invariance of $I_cR_n$
upon intercalation is strong evidence for an $s$-wave OP. \cite{kars}

Recently, Takano {\it et al.} crossed two single crystal Bi2212
whiskers an angle $\phi_0$
about the $c$-axis and fused them together,\cite{takano3,mos2002} using a technique
similar to that of Li {\it et al.}\cite{li}   In
addition to the nominal   composition of overdoped  Bi2212,
 with $T_c\approx 80$K, such
whiskers usually have a second transition at $T_{c2}\approx105$K, due to Bi$_2$Sr$_2$Ca$_2$Cu$_3$O$_{10+\delta}$ (Bi2223)
contamination.\cite{takano2,takanosjp}  For a 90$^{\circ}$
cross-whisker junction, a HRTEM picture revealed that this junction was uniform
over 100 nm, but the expected periodic lattice distortion was difficult to
discern.\cite{takanosjp} No HRTEM pictures of other cross-whisker junctions
were made.\cite{takanopc} 

Remarkably, Takano {\it et al.}  observed
branch structures in the current-voltage ($I-V$) characteristics of their
cross-whisker junctions at 5K, indicating that the insulating edge regions of
the two whiskers had somehow fused into a mesa structure,\cite{takano3}
consisting of a stack of Josephson junctions.\cite{yurgens,irie}  By
 assuming $I_c$ of the central $V=0$ branch corresponded to that of 
the twist junction and that
the junction area was equal to the entire whisker overlap, Takano
{\it et al.}  inferred a value for the junction  $J_c$ for
each of their 10-16 samples.\cite{takano3,mos2002}  In sharp contrast to the
 single crystal twist
experiments,\cite{li} $J_c(\phi_0)$ obtained in this way for
 the cross-whisker junctions
varied substantially with $\phi_0$.\cite{takano3,mos2002}

Nevertheless, for the 45$^{\circ}$ cross-whisker junction in a parallel magnetic
field, a  
 Fraunhofer-like diffraction pattern  consistent with the long-junction 
limit  was observed for their junction with  width 
36.7$\mu$m and  
Josephson length $\lambda_J\approx3-4\mu$m.\cite{mos2002,irie} In addition, Shapiro step
analysis of the 45$^{\circ}$ cross-whisker junction indicated that only 
first-order tunneling is present. \cite{tachiki}  If we assume those results are correct, then the data for the crucially important region
$\phi_0\approx45^{\circ}$ arises from {\it  weak, first-order tunneling} only. Since at 5K,
$I_c(45^{\circ})=0.227$mA          is orders of
magnitude larger than the minimum measureable $I_c$, \cite{mos2002}
 we use a logarithmic scale to fit the $J_c$  data. Since the overdoped
whiskers had $T_c\approx80$ K, \cite{mos2002} we take the maximum gap value in
 our fits to be $\Delta_0=22$meV, consistent with single crystal point contact
tunneling values.\cite{johnz}

Although  $J_c(45^{\circ})\ne0$, Takano {\it et al.}
nevertheless claimed that the strong, four-fold periodic $\phi_0$ dependence of $J_c(\phi_0)$ was evidence for
 a ``$d$-like'' OP.\cite{takano3,mos2002}  Here we show that
 such  $J_c(\phi_0)$ behavior, if correct, is
 merely a signal of coherent tunneling in a nearly tetragonal crystal with a
non-circular Fermi surface. Then, only the behavior of $J_c(\phi_0)$ for
 $\phi_0\approx45^{\circ}$, where the Josephson tunneling can safely be taken to be weak
and first order, is sensitive to the OP symmetry. Quantitative fits
 to the data of Takano {\it et al.} are obtained with either a very
anisotropic $s$-wave gap function  on the tightbinding Fermi surface, or a constant gap $\Delta_0$ limited to the extended Van Hove saddle
bands,\cite{vanhove} but the only  possibilities of a predominantly $d$-wave OP to
 fit the data are very unlikely.

 For weak tunneling across a cross-whisker
Josephson junction with cross-angle $\phi_0$, $J_c$ is given by \cite{bks}
\begin{eqnarray}
J_c(\phi_0)&=&\bigl|4eT\sum_{\omega}\langle f^J({\bf k},{\bf
 k}')F_{\omega}({\bf k})F^{*}_{\omega}(\tilde{{\bf k}'})\rangle\bigr|,
\end{eqnarray}
where we have set $\hbar=c=k_B=1$, $\langle\cdots\rangle$ represents the integrations over the first
 Brillouin zones (BZ's) on each side of the junction, $F_{\omega}({\bf k})=\Delta({\bf
 k},T)/[\omega^2+\xi^2({\bf k})+|\Delta({\bf k},T|^2]$ is the anomalous Green
 function, $\Delta({\bf k},T)$ is the OP with wavevector ${\bf k}$, $\xi({\bf
 k})$ is the  quasiparticle dispersion,  the $\omega$ are the Matsubara
 frequencies, and $\tilde{\bf k}$ represents the wavevectors ${\bf k}=(k_x,k_y)$ rotated by
 $\phi_0$ about the $c$-axis.\cite{bks}  Here we  set $T_c=80$ and $T=9$K,
 sufficiently close to 5K, allowing us to include only 500 Matsubara
 frequencies in the sum in Eq. (1). For overdoped cross whiskers,
 the normal state is rather metallic, and the BCS-like model, employed above, with general
 orbital OP symmetry, is
  a good approximation.
 
  For strongly incoherent tunneling, $J_c(\phi_0)/J_c(0)=1$  for all $s$-wave
OP forms, and although for all $d$-wave forms, $J_c(0)$ is vanishingly small,
$J_c(\phi_0)/J_c(0)\approx|\cos(2\phi_0)|$, regardless of the
details of the quasiparticle states.\cite{bks}  Neither form fits the
cross-whisker data. \cite{takano3,mos2002} However, for the
quasiparticle states pictured in Fig. 1, a substantial fraction
of coherent tunneling necessarily leads to a strong, four-fold $\phi_0$
dependence to $J_c(\phi_0)$ for $\phi_0\approx0^{\circ}, 90^{\circ}$,\cite{arnold,bks} even for an isotropic $s$-wave OP, 
and $J_c(\phi_0)$ is mainly sensitive to the OP symmetry for
$\phi_0\approx45^{\circ}$.\cite{arnold,bks} Regardless of the tunneling
coherence, for weak
tunneling between  single-component, nearly tetragonal $d$-wave 
superconductors,  
 at some particular $\phi_0$ value $\phi_0^*\approx45^{\circ}$, the
quantity inside the absolute value signs in Eq. (1) {\it changes sign}, and hence $J_c(\phi^*_0)=0$.\cite{bks,krs}  

 In order to fit the whisker data, we therefore assume coherent tunneling, $f^J({\bf
 k},{\bf k}')=f_0\delta^{(2)}({\bf k}-{\bf k}')$, and  take $\xi({\bf k})$ to have
 either  the tightbinding form, $\xi({\bf
 k})=-t[\cos(k_xa)+\cos(k_ya)]+t'\cos(k_xa)\cos(k_ya)-\mu$, where $t=306$ meV,
 $t'/t=0.90$, and $\mu/t=-0.675$, \cite{bks} or to represent the  extended
 `van
 Hove'  saddle
 bands, $\xi({\bf
 k})=-t[|\cos(k_xa)-\cos(k_ya)|-\nu]$, as shown in Fig. 1.  Here we take $t=500$ meV and
 $\nu=2.02$, so that the extended van Hove states never cross the Fermi energy $E_F$, but
 lie just below it in the vicinity of the $\overline{M}$ points at
 $(0,\pi/a)$, etc. This form is a rather crude approximation to the actual van
 Hove states, but serves to illustrate the effects of a cross-whisker junction
 rather well.  Since both of these quasiparticle state forms are
 periodic, umklapp processes are automatically included in our
 calculations.\cite{arnold}  For simplicity of notation, we set
 $\phi({\bf k})\equiv[\cos(k_xa)-\cos(k_ya)]N$, so the ordinary $d$-wave OP is
 $\Delta_0\phi({\bf k})$, where $N=0.5$ for the van Hove case, and
 $N=0.5315$  for the tight-binding case, which lead to the maximum value of
 the gap $\Delta_0=22$ meV at the respective closest approaches to the $\overline{M}$ points
 in the BZ.

\begin{figure}[t]
\includegraphics[width=0.45\textwidth]{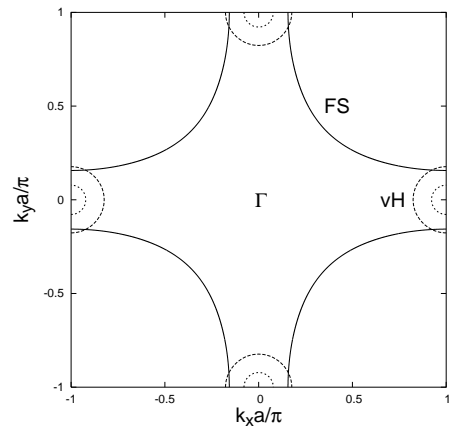}
\caption{Quasiparticle excitation features used in the fits.  Solid:
tightbinding Fermi surface. Dotted and dashed: Van Hove lines of constant
energy -25 meV and -90 meV, respectively, relative to $E_F$.
}
\label{fig1} 
\end{figure}

 A mixed $d+is$ OP could occur below a second phase transition, provided that
the $bc$ plane containing the periodic lattice distortion is indeed a good
mirror plane.\cite{krs}  However, if the disorder suggested by the STM
measurements on cleaved single crystal Bi2212 were  present in the Bi2212
whiskers,\cite{lang} then there might not be any relevant mirror plane, and
a mixed $d\pm s$ type of OP could occur without a second phase transition.
Hence, we  considered both OP forms. A
$d_{x^2-y^2}+id_{xy}$ OP  is
inconsistent with both the single crystal twist experiment and with the Pb(Nb)/Bi2212 
Josephson junction experiments,\cite{moessle,kawayama,li} with consequences
very similar to those of the $d_{x^2-y^2}+is$ state.

\begin{figure}[t]
\includegraphics[width=0.4\textwidth,angle=-90]{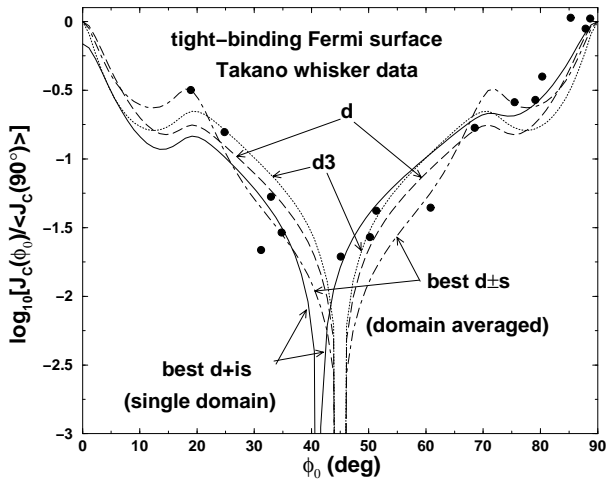}
\caption{Plot of $\log_{10}[J_c(\phi_0)/<J_c(90^{\circ})>]$ versus $\phi_0$
obtained from Ref. [9] (solid circles).
Also shown are the fits at 9K assuming coherent $c$-axis tunneling,  the
tight-binding $\xi({\bf k})$, and OP's of the ordinary $d$-wave,
 ($d$, long-dashed) and $d$-wave
cubed,  ($d3$, dotted) forms, along with
the best $d+is$ single domain fit,
$\epsilon=0.15$, (solid)
and the domain-averaged result of the best single domain $d\pm s$ fit,
 (dot-dashed).  See the text.}
\label{fig2} 
\end{figure}

 The $d+is$
scenario might at first sight appear to be consistent with 
the observation of an effect claimed to be due to 
spontaneous time-reversal symmetry breaking below the
pseudogap onset in underdoped Bi2212.\cite{kaminski}  However, in the
overdoped regime of the cross-junction
 experiments, the angle-resolved photoemission spectroscopy (ARPES) measurements were only made  above $T_c$, 
and the effect was not observed.\cite{kaminski}
There is very strong evidence that the pseudogap exists above $T_c$ for all
Bi2212 dopings, and that it is distinct from the superconducting gap.
 \cite{lang,krasnov,heim,shibauchi} Whatever the source of the effect, if it
 were simply a 
property of the pseudogap, it
should have been seen for all Bi2212 dopings. Since the  effect was only seen 
in the underdoped regime, where Bi2212 is known to be very
strongly disordered,\cite{lang} the effect itself might be a combined
property of the pseudogap and the disorder.  In any event, it is unlikely to
be relevant to the superconducting OP in the overdoped regime. 

In Fig. 2, we show our results for the best $d$-wave fits to the data,
assuming the tightbinding $\xi({\bf k})$ form.  In this and subsequent
figures, we assumed coherent $c$-axis tunneling, 
$T=9$ K and took the value  $\langle J_c(90^{\circ})\rangle$ to be the average of the
three data points in that vicinity.  We then fit the data to the ordinary $d_{x^2-y^2}$-wave
form, $\Delta_0\phi({\bf k})$, a cubed $d$-wave OP,
$\Delta_0\phi^3({\bf k})$, as suggested by  recent ARPES
measurements on underdoped (Pb,Bi)$_2$Sr$_2$CaCu$_2$O$_{8+\delta}$, \cite{borisenko}  a
single-domain time-reversal symmetry broken $d+is$ state,
$\Delta_0[\phi({\bf k})+i\epsilon]/(1+\epsilon^2)^{1/2}$, and  a single-domain $d\pm s$ state,
$\Delta_0[\phi({\bf k})\pm\epsilon]/(1+|\epsilon|)$.  For the
$d\pm s$ state, the best fit was for $\epsilon=0.25$.  Over the region of
available data, this  curve was
nearly indistinguishable from the $d+is$ curve with $\epsilon=0.15$ shown in
Fig. 2.    In Fig. 2, we also show the result of a multi-domain
average of the  $d+s$, $d-s$ domains with $\epsilon=0.25$, which includes the
identical contributions of $d\pm s|d\pm s$  domains across the
cross-whisker junction, and  the inequivalent $d\pm s|d\mp s$ domain
contributions. 

The best single domain $d+is$ and $d\pm s$  fits  straddled the dataless region near to 40$^{\circ}$, and hence
could be consistent with the data for $0.1\le\epsilon\le0.25$.  However,
 this time-reversal symmetry breaking
$d+is$ state has a vanishing $J_c(\phi_0)$ in the dataless regime in the
vicinity of $\phi_0=40^{\circ}$,
but not also at the crystallographically identical cross-whisker angle
$50^{\circ}$, for which nonvanishing data were available.  Hence, this state 
 would require a
single $d+is$ domain, in apparent contradiction with STM studies.\cite{lang}  
For a mixed $d+is$
or $d\pm s$ state with multiple domains, averaging over the domains always
leads  to $J_c(45^{\circ})=0$, as pictured in
Fig. 2.  We note that averaging over multiple $d+is$ domains with the
optimal $\epsilon=0.15$ results in a curve that is nearly indistinguishable
over the region of available data from the $d$-cubed state plotted in Fig. 2.
        
\begin{figure}[b]
\includegraphics[width=0.35\textwidth,angle=-90]{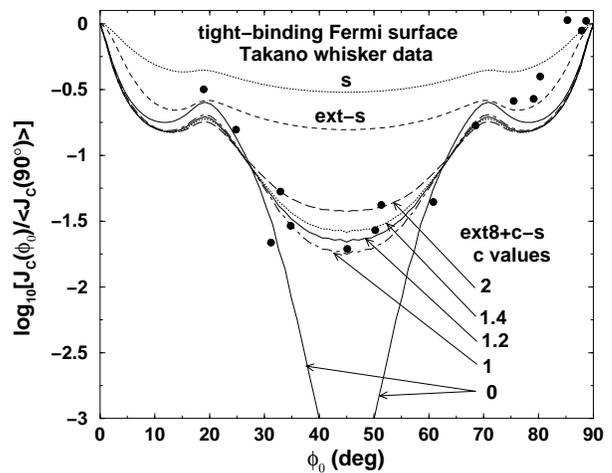}
\caption{Plot of $\log_{10}[J_c(\phi_0)/<J_c(90^{\circ})>]$ versus $\phi_0$
obtained from Ref. [9] (solid circles).
Also shown are the fits obtained at 9K with the
tight-binding $\xi({\bf k})$, and OP's of the  isotropic $s$-wave
(``s'', dotted) and extended-$s$-wave
 (``ext-s'', dashed) forms, and of the anisotropic ``ext8+c-s'',
$(\Delta_0-\Delta_c)\phi^8({\bf k})+\Delta_c$, forms with the$\Delta_c$ values in
meV of 2 (long-dashed), 1.4 (dotted), 1.2 (solid), 1, (dot-dashed), and 0
(solid), as indicated.}
\label{fig3} 
\end{figure}

In Fig. 3, we show the best $s$-wave fits using the same tightbinding
$\xi({\bf k})$.  Here we show fits to the isotropic $s$-wave OP,
$\Delta_0$, the extended-$s$-wave OP, $\Delta_0|\phi({\bf k})|$, and
several values of the highly anisotropic-$s$-wave OP,
$(\Delta_0-\Delta_c)\phi^8({\bf k})+\Delta_c$ (indicated in Fig. 3 by
`ext8+c-s') for $\Delta_c=0, 1, 1.2, 1.4,$ and 2 meV, as indicated. The best fits
are
for $\Delta_c\approx1.2-1.4$ meV, but $1\le\Delta_c\le2$ meV is
acceptable. These values are compatible with the resolution of  recent
ARPES experiments. \cite{borisenko} The flat ${\bf k}$ dispersion away from the minimum
gap position on the tightbinding Fermi surface is also compatible with the
ARPES data. We note that  ARPES
data are  complicated by the non-superconducting pseudogap, which
appears at $T^*>T_c$ for all Bi2212 dopings,\cite{krasnov,shibauchi} so that the superconducting gap
at each ${\bf k}$ of observation is merely constrained to be less than or equal to
the  gap observed by ARPES.  We note that the curves in Figs. 2 and 3 for the
ordinary $d$, ordinary $s$, and extended-$s$-wave OP's differ slightly from
those presented previously, \cite{bks} since those curves were calculated just
below $T_c$, and these results are for 9K$<<T_c$.

\begin{figure}[t]
\includegraphics[width=0.35\textwidth,angle=-90]{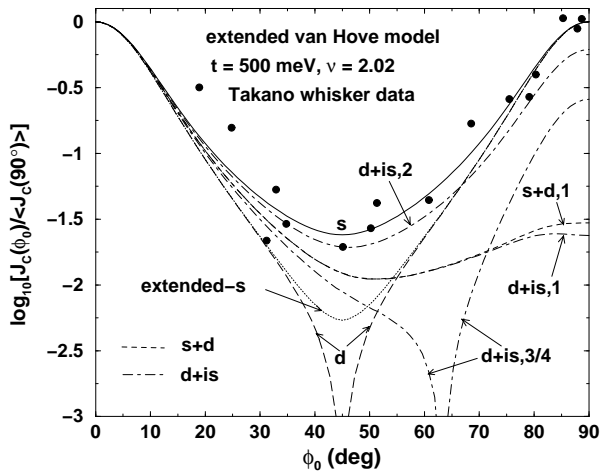}
\caption{Plot of $\log_{10}[J_c(\phi_0)/<J_c(90^{\circ})>]$ versus $\phi_0$
obtained from Ref. [9] (solid circles).
Also shown are the fits obtained from the
extended van Hove $\xi({\bf k})$, for OP's of the ordinary $d$-wave (long-dashed), ordinary $s$-wave 
(solid), extended-$s$-wave, 
 (dotted), and  the $d+is$ forms (dot-dashed) for the $\epsilon$
values indicated. For $\epsilon=1$, results for the $s+d$ state
(short-dashed) are also shown.}
\label{fig4} 
\end{figure}

Finally, we present our fits relevant to the van Hove scenario.  Here we
adjusted the bandwidth $t$ and the maximum of the saddle bands $-t\nu$
relative to $E_F$ to
obtain the best fit for  the ordinary $s$- and $d$-wave
OP's. 
The values shown here, $t=500$ meV and $\nu=2.02$, are intermediate to 
 both optimal values.  In addition, we showed the calculations for the
single domain $d+is$ (or $s+id$) state, $\Delta_0[\phi({\bf
k})+i\epsilon]/(1+\epsilon^2)^{1/2}$ for $\epsilon=0.75, 1,$ and 2, respectively.
For $\epsilon=1$, we also showed the results for the single domain $s\pm d$
state, $\Delta_0[\phi({\bf k})\pm\epsilon]/(1+|\epsilon|)$.  We note that the
ordinary $d$-wave curve and the 
three curves with $\epsilon=0.75, 1$ are inconsistent with the data, but the curve
with $\epsilon=2$ is consistent with the data. However, this (predominantly
$s$-wave) $d+is$ state exhibits a strong
amount of time-reversal symmetry breaking, and is hence unlikely to be
compatible with a variety of other experiments, as noted above.

We remark that the highly anisotropic OP ``ext8+c-s'' used
phenomenologically in Fig. 3 to fit the data could arise from a van Hove
scenario, in which the dominant pairing occurs over the van Hove bands
pictured in Fig. 1, and appears on the tight-binding Fermi surface by weak
coupling of the electronic states, as discussed elsewhere.
\cite{vanhove,tachiki2}  We note that a
good fit could be obtained within the van Hove scenario using an isotropic
$s$-wave OP, as shown in Fig. 4, so the physics of the
generalized $s$-wave OP's in Figs. 3 and 4 need not be
substantially different or exotic.

Since this work was submitted for publication, it came to our attention that 
 evidence exists that might cause one to suspect that some of the $J_c(\phi_0)$
 values  reported by Takano {\it et al.} might 
not represent the
 intrinsic $J_c(\phi_0)$ of the cross-whisker junctions.\cite{takanosjp} 
 For the same
$90^{\circ}$, $75^{\circ}$, and $60^{\circ}$ cross-whisker junctions for which
 the mesa branchings were shown, \cite{takano3} the measured resistances $R$ in the $T$ region
 $70-75$K$\approx T_c\le T\le 105$K  revealed
significant complications, suggestive of the presence of 
more superconducting Bi2223 at the 90$^{\circ}$ cross-whisker interface than
 for the 75$^{\circ}$ cross whisker, and 
what might be evidence for a non-superconducting or possibly even
 {\it insulating} 
barrier at the junction of the 60$^{\circ}$ cross whisker.\cite{takanosjp}
 Extensive studies of Bi2212 mesas cut from single crystals 
 established
 that the central $V=0$ branch, with the smallest  $I_c$,
 is frequently associated with the junction closest to the current lead, and
hence furthest from the mesa
 center.\cite{irie} In addition, the branch with the  lowest  $I_c$
corresponds to the true critical current of the stack.\cite{irie}  These
 properties might lead to an overestimate of $J_c(90^{\circ})$, and
 an underestimation of $J_c(60^{\circ})$, for example.
 Hence it is possible  that the intrinsic $\phi_0$ dependence of $J_c$ might
 be weaker than reported.

In summary, we have shown that the recent cross-whisker Josephson junction results of Takano
{\it et al.}, while different in detail from the single crystal results of Li
{\it et al.}, also render a predominantly $d$-wave OP form unlikely.
  As for a mixed $s$- and $d$-
wave OP, there is a narrow window of 10-25\% $s$-wave that would be allowed in
a particular single domain of mixed OPs, but otherwise, it appears that one is forced to
accept the result that the dominant OP is indeed $s$-wave, although it could
be highly anisotropic.  Although the results are compatible with an isotropic
OP on an  extended van Hove saddle band of quasiparticle states, if the
pairing were to take place mainly on the tightbinding Fermi surface, the
$s$-wave gap functions would have to be highly anisotropic, with a minimum
value in the range 1-2 meV, consistent with ARPES experiments.\cite{borisenko}

In order to strengthen these conclusions, we urge that additional data points
in the range $30^{\circ}\le\phi_0\le60^{\circ}$ be taken, and that the
temperature of the measurement be raised up near to $T_c$.\cite{krs}  
A few more junctions with $\phi_0\approx45^{\circ}$ are currently
under study, and the preliminary results appear to be consistent with the
above data.\cite{takanopc}
We would also like to see  other measurements to investigate if the
quasiparticle $c$-axis transport is indeed coherent, unlike Bi2212 single
crystals.   We also
 urge that resistivity measurements between $T_c$ and 105K be made on 
cross-whisker junctions with $\phi_0\approx45^{\circ}$, and that alternative fabrication procedures be
 investigated, in order to guarantee their uniformity.

The author would like to thank  K. Kadowaki, R. Kleiner, Q. Li,
 K. Scharnberg, J. R. Schrieffer,  J. Sethna, and  M. Tachiki
for useful discussions,
  and
especially Y. Takano for a preprint containing his latest data, a copy
of Ref. [9], and extensive communications.   

{\it Note to be added in proof:}  It has ultimately come to our attention that an attempt to fit 
the whisker data of
Takano {\it et al.} was made by Maki and Haas.\cite{maki} There are three
problems with that work.  First, those authors
assumed the Fermi surface of Bi2212 to be a cylinder with a circular
cross-section centered about the central $\Gamma$ point in the first
BZ, inconsistent with ARPES experiments on Bi2212.\cite{borisenko}  Second, they
 ascribed the $\phi_0$ dependence of $J_c$ to Andreev scattering at the 
junction interfaces, which they assumed to be between superconducting  and
normal metallic  layers,\cite{maki} inconsistent with the 
Shapiro step analysis and
the multiple branching behavior of the whisker current-voltage
characteristics, which provide
compelling evidence that the junctions are weak, first order, and of the
superconducting-insulating variety, \cite{takano3,tachiki} as generically
 occurs in single crystal
Bi2212.\cite{irie,krasnov} Third, their formulas (3) and (6) are
inconsistent with those obtained for their model from the exact expression,
Eq. (18) of Ref. \cite{arnold},
obtained for the $c$-axis critical current for coherent tunneling between two
layered superconductors.  Hence, the $J_c(\phi_0)$ obtained by Maki and 
Haas is both mathematically imprecise and physically
very unlikely to apply to Bi2212 whiskers.

\end{document}